%% file: main.tex
\documentclass[11pt]{article}

\usepackage{fullpage}
\usepackage[utf8]{inputenc}
\usepackage[english]{babel}
\usepackage{soul}

\usepackage[colorinlistoftodos]{todonotes}

\usepackage{calc}
\usepackage{tikz}
\usepackage{enumitem}
\usepackage{amsthm}
\usepackage{amsmath}
\usepackage{array}

\usepackage{amssymb}
\usepackage{comment}
\usepackage{latexsym}
\usepackage{xspace}
\usepackage{hyperref}

\newtheorem{theorem}{Theorem}
\newtheorem{lemma}[theorem]{Lemma}
\newtheorem{corollary}[theorem]{Corollary}

\newtheorem{remark}[theorem]{Remark}

\input defs

\begin{document}

\title{Parity Decision Tree Complexity is Greater Than Granularity
}

\author{Anastasiya Chistopolskaya\thanks{National Research University Higher School of Economics} \and Vladimir V. Podolskii\thanks{Steklov Mathematical Institute, Russian Academy of Sciences and
National Research University Higher School of Economics}
}

\date{}

\large

\maketitle
\input body.tex

\end{document}

%% file: defs.tex
\newcommand{\moo}{\{-1,1\}}

\newcommand{\zoo}{\{0,1\}}
\newcommand{\zoon}{\{0,1\}^n}
\newcommand{\bb}[1]{\mathbb{#1}}
\newcommand{\wh}[1]{\widehat{#1}}

\newcommand{\Supp}{\mathsf{Supp}}
\newcommand{\spar}{\mathsf{spar}}
\newcommand{\gran}{\mathsf{gran}}
\newcommand{\rank}{\mathrm{rank}}

\newcommand{\pow}{\mathsf{P}}
\newcommand{\bin}{\mathsf{B}}

\DeclareMathOperator*{\Exp}{\mathbf{E}}

\newcommand{\dcc}{\mathsf{D}^{\mathsf{cc}}}
\newcommand{\pdec}{\mathsf{D}_{\oplus}}
\newcommand{\cert}{\mathsf{C}_{\oplus}}
\newcommand{\mult}{\mathsf{c}_{\wedge}}

\newcommand{\cir}{\mathcal{C}}

\newcommand{\la}{\langle}
\newcommand{\ra}{\rangle}

\newcommand{\MAJ}{\ensuremath{\operatorname{MAJ}}}
\newcommand{\AND}{\ensuremath{\operatorname{AND}}}
\newcommand{\NOT}{\ensuremath{\operatorname{NOT}}}
\newcommand{\RMAJ}[1]{\ensuremath{\operatorname{MAJ}}^{\otimes #1}_3}
\newcommand{\THR}{\ensuremath{\operatorname{THR}}}

\newcommand{\IP}{\ensuremath{\operatorname{IP}}}

%% file: body.tex

\begin{abstract}
We prove a new lower bound on the parity decision tree complexity $\mathsf{D}_{\oplus}(f)$ of a Boolean function $f$. Namely, granularity of the Boolean function $f$ is the smallest $k$ such that all Fourier coefficients of $f$ are integer multiples of $1/2^k$. We show that $\mathsf{D}_{\oplus}(f)\geq k+1$.

This lower bound is an improvement of lower bounds through the sparsity of $f$ and through the degree of $f$ over $\mathbb{F}_2$. Using our lower bound we determine the exact parity decision tree complexity of several important Boolean functions including majority and recursive majority. For majority the complexity is $n - \mathsf{B}(n)+1$, where $\mathsf{B}(n)$ is the number of ones in the binary representation of $n$. For recursive majority the complexity is $\frac{n+1}{2}$. Finally, we provide an example of a function for which our lower bound is not tight.

Our results imply new lower bound of $n - \mathsf{B}(n)$ on the multiplicative complexity of majority.
\end{abstract}


\section{Introduction}

Parity decision trees is a computational model in which we compute a known Boolean function $f\colon \zoon \to \moo$ on an unknown input $x \in \zoon$ and in one query we can check the parity of arbitrary subset of inputs. The computational cost in this model is the number of queries we have made. The model is a natural generalization of a well-known decision trees model (in which only the value of a variable can be asked in one query)~\cite{BuhrmanW02,Jukna12}. 

Apart from being natural and interesting on its own parity decision trees model was studied mainly in connection with Communication Complexity and more specifically, with Log-rank Conjecture. In Communication Complexity most standard model there are two players Alice and Bob. Alice is given $x \in \zoon$ and Bob is given $y \in \zoon$ and they are trying to compute some fixed function $F \colon \zoon \times \zoon \to \moo$ on input $(x,y)$. The question is how many communication is needed to compute $F(x,y)$ in the worst case. It is known that the deterministic communication complexity $\dcc(F)$ of the function $F$ is lower bounded by $\log \rank(M_F)$, where $M_F$ is a communication matrix of $F$~\cite{KushilevitzN97}. It is a long standing conjecture and one of the key open problems in Communication Complexity, called Log-rank Conjecture~\cite{LovaszS88}, to prove that $\dcc(F)$ is upper bounded by a polynomial of $\log \rank(M_F)$. 

An important special case of Log-rank Conjecture addresses the case of XOR-functions $F(x,y) = f(x\oplus y)$ for some $f$, where $x\oplus y$ is a bit-wise XOR of Boolean vectors $x$ and $y$. On one hand, this class of functions is wide and captures many important functions (including equality, inner product, Hamming distance), and on the other hand the structure of XOR-functions allows to use analytic tools. For such functions $\rank(M_F)$ is equal to the Fourier sparsity $\spar f$, the number of non-zero Fourier coefficients of $f$. Thus, the Log-rank Conjecture for XOR-functions can be restated: is it true that $\dcc(F)$ is bounded by a polynomial of $\log \spar f$?

Given a XOR-function $f(x\oplus y)$ a natural way for Alice and Bob to compute the value of the function is to use a parity decision tree for $f$. They can simulate each query in the tree by computing parity of bits in their parts of the input separately and sending the results to each other. One query requires two bits of communication and thus $\dcc(F) \leq 2\pdec(f)$. This leads to an approach to establish Log-rank Conjecture for XOR-function~\cite{ZhangS10}: show that $\pdec(f)$ is bounded by a polynomial of $\log \spar f$.

This approach received a lot of attention in recent years and drew attention to parity decision trees themselves~\cite{ZhangS10,TsangWXZ13,ShpilkaTV14,Yao16,TsangXZ16,HatamiHL16}. In a recent paper~\cite{HatamiHL16} it was shown that actually $\dcc(F)$ and $\pdec(f)$ are polynomially related. This means that the simple protocol described above is not far from being optimal and that the parity decision tree version of Log-rank Conjecture stated above is actually equivalent to the original Log-rank Conjecture for XOR-functions.

All this motivates further research on parity decision trees. As for the lower bounds for parity decision tree complexity, one technique follows from the discussion above: $\pdec(f) \geq \dcc(F)/2\geq (\log \spar f)/2$. Although, if Log-rank conjecture for XOR-functions is true, this approach gives optimal bounds up to a polynomial, in many cases it does not help to determine the precise parity decision tree complexity of Boolean functions. For example, this approach always gives bounds of at most $n/2$ for functions of $n$ variables.

Another known approach is of a more combinatorial flavor. For standard decision trees there are several combinatorial measures known that lower bound decision tree complexity. Among them the most common are certificate complexity and block sensitivity. In~\cite{ZhangS10} these measures were generalized to the setting of parity decision tree complexity. Parity decision tree complexity versions of these measures are actually known to be polynomially related to parity decision tree complexity~\cite{ZhangS10}. However, they also do not give tight lower bounds for some interesting functions.

Yet another standard approach is through the degree of polynomials. It is well known that the complexity of a function in standard decision trees model is lower bounded by the degree of the function over $\bb{R}$ (see, e.g.~\cite{BuhrmanW02}). Completely analogously it can be shown that the parity decision tree complexity of a function is lower bounded by the degree of the function over $\bb{F}_2$ (although, the adaptation to parity decision trees is straightforward we have not seen it mentioned in the literature; we provide a proof in Section~\ref{sec:prelim} for the sake of completeness). This approach also does not give tight lower bounds for some interesting functions.

Examples of well-known functions for which the precise parity decision trees complexity is unknown include the majority function (playing a crucial role in many areas of Theoretical Computer Science, including Fourier analysis of Boolean functions) and recursive majority (interesting, in particular, from decision tree complexity point of view as it provides a gap for deterministic and randomized decision tree complexity~\cite{SaksW86,MagniezNSSTX16}).

\paragraph*{Our Results} In this paper we address the problem of improving known lower bounds for parity decision tree complexity. Our main result is a new lower bound in terms of the \emph{granularity} of a Boolean function.

Granularity $\gran(f)$ of $f \colon \zoon \to \moo$ is the smallest $k$ such that all Fourier coefficients of $f$ are integer multiples of $1/2^k$. We show that 
$$
\pdec(f) \geq \gran(f)+1.
$$

It is a simple corollary of Parseval's Identity that $\gran(f) \geq (\log\spar (f))/2$. Thus our lower bound is an improvement over the bound through sparsity. On the other hand, it was shown in~\cite{GopalanOSSW11} (see also~\cite{TsangXZ16}) that $\gran(f) \leq \log \spar(f)-1$. Thus, this is an improvement by at most a factor of $2$. 

We also observe that $\gran(f) +1 \geq \deg_2(f)$, where by $\deg_2(f)$ we denote the degree of $f$ over $\bb{F}_2$. Thus, our lower bound is also not weaker than the lower bound through the degree of the function. 

Despite for our lower bound being close to the lower bound through sparsity, it allows to prove tight lower bounds for several important functions. Also unlike the lower bound through sparsity, new approach allows to prove lower bounds up to $n$ (the largest possible parity decision tree complexity of a function). 

We hope that the connection between parity decision tree complexity and granularity will help to shed more light on the parity decision tree complexity.

We apply our lower bound to study the parity decision tree complexity of several well-known Boolean functions. We start with the majority function $\MAJ$. We show that $\pdec(MAJ) = n - \bin(n)+1$, where $n$ is the number of variables and $\bin(n)$ is the number of ones in the binary representation of $n$. The upper bound in this result is a simple adaptation of a folklore algorithm for the following problem (see, e.g.~\cite{SaksW91}). Suppose that for odd $n$ we are given $n$ balls of red and blue colors and we do not see the colors of the balls. In one query for any pair of balls we can check whether their colors are the same. Our goal is to find a ball of the same color as the majority of balls. We want to minimize the number of queries asked in the worst case. There is a folklore algorithm to solve this task in $n- \bin(n)$ queries. It was shown in~\cite{SaksW91} that this is in fact optimal. On the idea level our lower bound for parity decision tree complexity is inspired by the proof of~\cite{SaksW91}.

Due to the connection between parity decision tree complexity and multiplicative complexity communicated to us by Alexander Kulikov~\cite{Kulikov2018} from our results it follows that multiplicative complexity of $\MAJ$ is at least $n - \bin(n)$. This is an improvement of the lower bound of~\cite{BoyarP08}. Previously our lower bound was known only in the case when $n$ is the power of $2$~\cite{BoyarP08}.

Next we proceed to recursive majority that computes an iteration of majority of three variables. We show that the parity decision tree complexity of this function is $(n+1)/2$. 

Finally, we show a series of examples of functions, for which our lower bound is not optimal. Namely, we consider threshold functions $\THR_n^l$ that check whether there are at least $l$ ones in the input. We show that for $n=8k+2$ for $k>0$ and $l=3$ our lower bound implies that at least $n-2$ queries are needed to compute the function, whereas the actual parity decision tree complexity is $n-1$. To prove this gap we combine our lower bound with an additional inductive argument allowing for a weak form of hardness amplification for the parity decision tree complexity of $\THR_n^k$ functions. 
 
The rest of the paper is organized as follows. In Section~\ref{sec:prelim} we provide necessary definition and preliminary information. In Section~\ref{sec:main} we prove the lower bound on parity decision tree complexity. In Sections~\ref{sec:maj}~and~\ref{sec:rmaj} we study the parity decision tree complexity of majority and recursive majority respectively. Finally, in Section~\ref{sec:example} we provide an example of a function for which our lower bound is not tight. Some of the technical proofs are moved to Appendix.

\section{Preliminaries} \label{sec:prelim}

\subsection{Fourier Analysis}

Throughout the paper we assume that Boolean functions are functions of the form $f \colon \zoon \to \moo$. That is, input bits are treated as 0 and 1 and to them we will usually apply operations over $\bb{F}_2$. Output bits are treated as $-1$ and $1$ and the arithmetic will be over $\bb{R}$. The value $-1$ correspond to `true' and $1$ corresponds to `false'.

We denote the variables of functions by $x=(x_1,\ldots,x_n)$. We use the notation $[n]=\{1,\ldots, n\}$.

We briefly review the notation and needed facts from Boolean Fourier analysis. For extensive introduction see~\cite{O'Donnell14book}.

For functions $f,g \colon \zoon \to \bb{R}$ consider an inner product
$$
\la f,g\ra = \Exp_x f(x)g(x),
$$
where the expectation is taken over uniform distribution of $x$ on $\zoon$.

For a subset $S \subseteq [n]$ we denote by $\chi_{S}(x) = \prod_{i\in S} (-1)^{x_i}$ the Fourier character corresponding to $S$. We denote by $\wh{f}(S)=\la f,\chi_S \ra$ the corresponding Fourier coefficient of $f$.

It is well-known that for any $x\in \zoon$ we have $f(x) = \sum_{S\subseteq [n]} \wh{f}(S)\chi_S(x)$.

If $f\colon\zoon \to \moo$ (that is, if $f$ is Boolean) then the well-known Parseval's Identity holds:
$$
\sum_{S\subset [n]} \wh{f}^2(S) =1.
$$

By the \emph{support} of the Boolean function $f$ we denote
$$
\Supp(f) = \{S\subseteq [n] \mid \wh{f}(S)\neq 0\}.
$$
The \emph{sparsity} of $f$ is $\spar(f)=|\Supp(f)|$. Basically, the sparsity of $f$ is the $l_0$-norm of the vector of its Fourier coefficients.

Consider a binary fraction $\alpha$, that is $\alpha$ is a rational number that can be written in a form that its denominator is a power of 2. By the \emph{granularity} $\gran(\alpha)$ of $\alpha$ we denote the minimal integer $k\geq 0$ such that $\alpha \cdot 2^k$ is an integer.

We will also frequently use the following closely related notation. For an integer $L$ denote by $\pow(L)$ the maximal power of 2 that divides $L$. It is convenient to set $\pow(0)=\infty$.

Note that for Boolean $f$ the Fourier coefficients of $f$ are binary fractions. By the \emph{granularity} of $f$ we call the following value
$$
\gran(f) = \max_{S\subseteq [n]} \gran(\wh{f}(S)).
$$
It is easy to see that for any $f\colon \zoon \to \moo$ it is true that 
$$
0 \leq \gran(f) \leq n-1
$$
and both of these bounds are achievable (for example, for $f(x)=\bigoplus_i x_i$ and $f(x)=\bigwedge_i x_i$ respectively).

It is known that $\gran(f)$ is always not far from the logarithm of $\spar(f)$:
$$
\frac{\log \spar(f)}{2} \leq \gran(f) \leq \log \spar(f)-1.
$$
The first inequality can be easily obtained from Parseval's identity. The second is a non-trivial result implicit in~\cite[Theorem 3.3 for $\mu=0$]{GopalanOSSW11} (see also~\cite{TsangXZ16}). Again, both inequalities are tight (the first one is tight for inner product $\IP(x,y)=\bigoplus_i (x_i\wedge y_i)$ or any other bent function~\cite{O'Donnell14book}; the second one is tight for example for $f(x)=\bigoplus_i x_i$).

For a Boolean function $f\zoon \to \moo$ denote by $\deg_2(f)$ the degree of the multilinear polynomial $p \in \bb{F}_2[x_1,\ldots,x_n]$ computing $f$ as a Boolean function, that is  for all $x\in \bb{F}_2^n$ we have $p(x)=1$ if $f(x)=-1$ and $p(x)=0$ otherwise. It is well known that such multilinear polynomial $p$ is unique for any $f$ and thus $\deg_2(f)$ is well defined.

It is known that $\deg_2(f)\leq \log \spar(f)$ for any $f$~\cite{BernasconiC99}. We observe that the granularity is also lower bounded by the degree of the function.

\begin{lemma}
For any $f \colon \zoon \to \moo$ we have $\deg_2(f) \leq \gran(f)+1$.
\end{lemma}

\begin{proof}
The proof strategy is similar to the one of~\cite{BernasconiC99}.

For a function $f \colon \zoon \to \moo$ consider two subfunctions $f_0$ and $f_1$ on $n-1$ varaibles obtained from $f$ by setting variable $x_n$ to 0 and to 1 respectively. Note that for any $S\subseteq [n-1]$ we have
\begin{align*}
\wh{f}(S) &= \Exp_{x\in \zoon} f(x) \chi_S(x) \\ &= \frac 12 \Exp_{x \in \zoo^{n-1}} f_0(x) \chi_S(x) + \frac 12 \Exp_{x \in \zoo^{n-1}} f_1(x) \chi_S(x) = \frac 12 \wh{f}_0(S) + \frac 12 \wh{f}_1(S)
\end{align*}
and
\begin{align*}
\wh{f}(S\cup\{n\}) &= \Exp_{x\in \zoon} f(x) \chi_S(x)\\ &= \frac 12 \Exp_{x \in \zoo^{n-1}} f_0(x) \chi_S(x) - \frac 12 \Exp_{x \in \zoo^{n-1}} f_1(x) \chi_S(x) = \frac 12 \wh{f}_0(S) - \frac 12 \wh{f}_1(S).
\end{align*}
Thus, 
$$
\wh{f}_0(S) = \wh{f}(S) + \wh{f}(S\cup\{n\})
$$
and
$$
\wh{f}_1(S) = \wh{f}(S) - \wh{f}(S\cup\{n\}).
$$
In particular, the granularity of both $f_0$ and $f_1$ is not larger than the granularity of $f$. From this we conclude that the granularity of a subfunction of $f$ is at most the granularity of $f$.

Denote $d=\deg_2(f)$ and consider a monomial of degree $d$ in the polynomial $p$ for $f$. For simplicity of notation assume that this is the monomial $x_1\ldots x_d$. Fix all variables $x_i$ for $i>d$ to $0$. We get a subfunction $g$ of $f$ of $d$ variables and degree $d$. As discussed above $\gran(g)\leq \gran(f)$, so it is enough to show that $d \leq \gran(g)+1$. For this note that since the function $g$ is of maximal degree we have that $|g^{-1}(-1)|$ is odd (see, e.g.~\cite[Section 2.1]{Jukna12}). Thus,
$$
\wh{g}(\emptyset) = \Exp_{x\in \zoo^d} g(x) = \frac{1}{2^d} \left(|g^{-1}(1)| - |g^{-1}(-1)| \right) = \frac{1}{2^d} \left(2^n - 2|g^{-1}(-1)| \right)
$$
and the granularity of $\wh{g}(\emptyset)$ is $d-1$.
\end{proof}

\subsection{Parity Decision Trees}

A parity decision tree $T$ is a rooted directed binary tree. Each of its leaves is labeled by $-1$ or 1, each internal vertex $v$ is labeled by a parity function $L_v(x)=\bigoplus_{i \in S_v} x_i$ for some subset $S_v \subseteq [n]$. Each internal node has two outgoing edges, one labeled by $-1$ and another by 1. A computation of $T$ on input $x \in \zoon$ is the path from the root to one of the leaves that in each of the internal vertices $v$ follows the edge, that has label equal to the value of $\bigoplus_{i \in S_v} x_i$. Label of the leaf that is reached by the path is the output of the computation. The tree $T$ \emph{computes} the function $f \colon \zoon\to \moo$ iff on each input $x \in \zoon$ the output of $T$ is equal to $f(x)$. Parity decision tree complexity of $f$ is the minimal depth of a tree computing $f$. We denote this value by $\pdec(f)$.

One known way to lower bound parity decision tree complexity goes through communication complexity of XOR functions. We state the bound in the following lemma (see, e.g.~\cite{HatamiHL16}).
\begin{lemma} \label{lem:spar_lower_bound}
For any function $f\colon\zoon \to \moo$ we have
$$
\pdec(f) \geq \frac{\log \spar(f)}{2}.
$$
\end{lemma}
This lower bound turns out to be useful in many cases, especially when we are interested in the complexity up to a multiplicative constant or up to a polynomial factor. However, it does not always help to find an exact value of the complexity of the function and in principle cannot give lower bounds greater than $n/2$.

Another more combinatorial approach goes through analogs of certificate complexity and block sensitivity for parity decision trees~\cite{ZhangS10}. Since parity block sensitivity is always less or equal then parity certificate complexity and we are interested in lower bounds, we will introduce only certificate complexity here.

For a function $f\colon \zoon\to \moo$ and $x\in\zoon$ denote by $C_{\oplus}(f,x)$ the minimal co-dimension of an affine subspace in $\zoon$ that contains $x$ and on which $f$ is constant. The \emph{parity certificate complexity} of $f$ is $C_{\oplus}(f) = \max_x C_{\oplus} (f,x)$.

\begin{lemma}[\cite{ZhangS10}]
For any function $f\colon\zoon \to \moo$ we have
$$
\pdec(f) \geq C_{\oplus}(f).
$$
\end{lemma}
This approach allows to show strong lower bounds for some functions. For example, it can be used to show that $\pdec(\AND_n)=n$. However, for more complicated functions like majority or recursive majority this lemma does not give tight lower bounds.

Yet another approach to lower bounds for parity decision tree complexity is through polynomials. Although it is very similar to analogous connection for standard decision trees, we have not observed it in the literature.

\begin{lemma}
For any $f\colon \zoon \to \moo$ we have $\pdec(f) \geq \deg_2(f)$.
\end{lemma}

\begin{proof}
The proof of this lemma follows closely the proof connecting standard decision tree complexity of a function with its degree over $\bb{R}$ (see, e.g.~\cite{BuhrmanW02}).

Consider a parity decision tree $T$ computing $f$ with depth equal to $\pdec(f)$. Consider arbitrary leaf $l$ of this tree and consider the path in $T$ leading from the root to $l$. For computation to follow this path on input $x$ in each internal vertex $v$ the input $x$ must satisfy some linear restriction $L(x)=1$ ($L(x)$ is the parity $L_v(x)$ labeling $v$ if the path follows the edge labeled by $-1$ out of $v$ and $L(x)=L_v(x)\oplus 1$ if the path follows the edge labeled by $1$). Denote all these linear forms in these restrictions along the path by $L_1(x),\ldots, L_p(x)$, where $p\leq \pdec(f)$. Thus, on input $x$ we follow the path to $l$ iff $L_1(x) \wedge \ldots \wedge L_p(x)$ is satisfied. Denote this expression by $T_{l}(x)$.

Denote by $S$ the set of all leaves of $T$ that are labeled by $-1$. For any input $x$ we have that $f(x)=-1$ iff the computation path in $T$ reaches a leaf labeled with $-1$ iff 
$$
\bigoplus_{l \in S} T_l(x)=1.
$$
It is left to observe that the latter expression is a multilinear polynomial over $\bb{F}_2$ of degree at most $\pdec(f)$.
\end{proof}

\subsection{Multiplicative Complexity}

Multiplicative complexity $\mult(f)$ of a Boolean function $f$ is the minimal number of $\AND$-gates in a circuit computing $f$ and consisting of $AND$, $\oplus$ and $\NOT$ gates, each gate of fan-in at most 2 (for formal definitions from circuit complexity see, e.g.~\cite{Jukna12}). This measure was studied in Circuit Compexity~\cite{BoyarP08,KojevnikovK10,BoyarF14} as well as in connection to Cryptography~\cite{KolesnikovS08,Vaikuntanathan11} and providing an explicit function $f$ on $n$ variables with $\mult(f)>n$ is an important open problem.

The following lemma was communicated to us by Alexander Kulikov~\cite{Kulikov2018} and with his permission we include it with a proof.

\begin{lemma} \label{lem:mc_vs_pdec}
For any $f$ on $n$ variables 
$$
\pdec(f) \leq \mult(f) +1.
$$
\end{lemma}

\begin{proof}
The proof is by induction on $s=\mult(f)$. 

If $s=0$, then $f$ is computed by a circuit consisting of $\oplus$ and $\NOT$ gates and thus  $f$ is a linear form of its variables. We can compute it by one query in parity decision tree model.

For the step of induction, consider an arbitrary $f$ and consider a circuit $\cir$ computing  $f$ with the number of $\AND$-gates equal to $\mult(f)$. Consider the first $\AND$-gate $g$ in $\cir$. Both of its inputs compute linear forms over $\bb{F}_2$. Our decision tree algorithm queries one of inputs of $g$. Depending on the answer to the query, $g$ computes either constant 0, or its second input. In both cases the gate $g$ computes a linear form over $\bb{F}_2$, so we can simplify the circuit and obtain a new circuit $\cir'$ computing the same function on inputs consistent with the answer to the first query and with at most $s-1$ $\AND$-gates. By induction hypothesis in both cases the function computed by $\cir'$ is computable in parity decision tree model with at most $s$ queries. Overall, we make $s+1$ queries.
\end{proof}

\section{Lower Bound on Parity Decision Trees} \label{sec:main}

Through the connection to communication complexity it is known that $\pdec(f)\geq \frac{\log \spar(f)}{2}$ for any $f$. In our main result we improve this bound.

\begin{theorem} \label{thm:main}
For any non-constant $f \colon \zoon \to \moo$ we have
$$
\pdec(f)\geq \gran(f) + 1.
$$
\end{theorem}

\begin{proof}
We prove the theorem by an adversary argument. That is, we will describe the strategy for the adversary to answer queries of a parity decision tree in order to make the tree to make many queries to compute the output.

Denote $k=\gran(f)$ and denote by $S\subseteq [n]$ the subset on which the granularity is achieved, that is $k=\gran(\wh{f}(S))$. We have that
\begin{align*}
\wh{f}(S) &= \frac{1}{2^n}\sum_{x\in \zoon} f(x)\chi_{S}(x) = \frac{1}{2^n} \left(\sum_{x \in f^{-1}(1)} \chi_S(x) - \sum_{x \in f^{-1}(-1)} \chi_S(x)\right)\\
& = \frac{1}{2^n} \left(\sum_{x \in \zoon} \chi_S(x) - 2\cdot\sum_{x \in f^{-1}(-1)} \chi_S(x)\right).
\end{align*}
Note that the first sum in the last expression is equal to $2^n$ if $S=\emptyset$ and is equal to $0$ otherwise. Thus for the granularity of $\wh{f}(S)$ to be equal to $k$ the sum $\sum_{x \in f^{-1}(-1)}\chi_S(x)$ should be divisible by $2^{n-k-1}$ and should not be divisible by $2^{n-k}$. In other words (recall that $\pow(L)$ is the maximal power of $2$ that divides $L$), 
\begin{equation} \label{eq:main1}
\pow\left(\sum_{x \in f^{-1}(-1)}\chi_S(x)\right) = n-k-1.
\end{equation}

After each step of the computation the query fixes some parity of inputs to be equal to some fixed value. Denote by $C_i\subseteq [n]$ the set of inputs that are still consistent with the current node of a tree after step $i$, and on which the function is equal to $-1$. We have that $C_0=f^{-1}(-1)$. 

We will show that we can answer the queries in such a way that
\begin{equation} \label{eq:main2}
\pow\left(\sum_{x \in C_{i+1}}\chi_S(x)\right) \leq \pow\left(\sum_{x \in C_{i}}\chi_S(x)\right). 
\end{equation}
To see this observe that the $(i+1)$-st query splits the current set $C_i$ into two disjoint subsets $A$ and $B$. In particular,
$$
\sum_{x \in C_{i}}\chi_S(x) = \sum_{x \in A}\chi_S(x) + \sum_{x \in B}\chi_S(x).
$$
If both sums in the right-hand side are divisible by some power of $2$, then the left-hand side also is. Thus,
$$
 \min\left(\pow\left(\sum_{x \in A}\chi_S(x)\right), \pow\left(\sum_{x \in B}\chi_S(x)\right)\right) \leq \pow\left(\sum_{x \in C_{i}}\chi_S(x)\right).
$$
Pick for $C_{i+1}$ the set, on which the minimum in the left-hand side is achieved.

Suppose the protocol makes $t$ queries. The set of inputs that reach the leaf forms an affine subspace of Boolean cube of dimension at least $n-t$, on which the function $f$ must be constant. Thus the sum
$$
\sum_{x \in C_t}\chi_S(x)
$$
is the sum of a character over an affine subspace, and thus is equal to either 0, or $2^{n-t}$. In both cases
\begin{equation} \label{eq:main3}
\pow\left(\sum_{x \in A}\chi_S(x)\right) \geq n-t.
\end{equation}
Combining \eqref{eq:main1}-\eqref{eq:main3} we get
$$
n-k-1 \geq n-t
$$
and the theorem follows.
\end{proof}

\section{Majority Function} \label{sec:maj}

In this section we analyze parity decision tree complexity of the majority function $\MAJ_n \colon \zoon \to \moo$. The function is defined as follows:
$$
\MAJ_n(x) = -1 \Leftrightarrow \sum_{i=1}^{n} x_i \geq \frac n2.
$$

To state our results we will need the following notation: let $\bin(k)$ be the number of ones in a binary representation of $k$.

We start with an upper bound. The following lemma is a simple adaptation of the folklore algorithm (see, e.g.~\cite{SaksW91}).

\begin{lemma} \label{lem:maj_upper_bound}
$$
\pdec(\MAJ_n) \leq n - \bin(n)+1.
$$
\end{lemma}

\begin{proof}
Our parity decision tree will mostly make queries of the form $y\oplus z$ for a pair of variables. Note that such a query basically checks whether $y$ and $z$ are equal.

Our algorithm will maintain splitting of input variables into blocks of two types. We will maintain the following properties:
\begin{itemize}
\item the size of each block is a power of 2;
\item all variables in each block of type 1 are equal;
\item blocks of type 2 are balanced, that is they have equal number of ones and zeros.
\end{itemize}
In the beginning of the computation each variable forms a separate block of size one. During each step the algorithm will merge two blocks into a new one. Thus, after $k$ steps the number of blocks is $n-k$.

The algorithms works as follows. On each step we pick two blocks of type 1 of equal size. We pick one variable from each block and query the parity of these two variables. If the variables are equal, we merge the blocks into a new block of type 1. If the variables are not equal, the new block is of type 2. The process stops when there are no blocks of type 1 of equal size.

It is easy to see that all of the properties listed above are maintained. In the end of the process we have some blocks of the second type (possibly none of them) and some blocks of the first type (possibly none of them) of pairwise non-equal size. Note that the value of the majority function is determined by the value of variables in the largest block of type 1. Indeed, all blocks of type 2 are balanced and the largest block of type 1 has more variables then all other blocks of type 1 in total. Thus, to find the value of $\MAJ_n$ it remains to query one variable from the largest block of type 1. Note, that the case when there are no blocks of type 1 in the end of the process correspond to balanced input (and even $n$). In this case we can tell that the output is $-1$ without any additional queries.

Note that the sum of sizes of all blocks is equal to $n$. Since the size of each block is a power of $2$, there are at least $\bin(n)$ blocks in the end of the computation (one cannot break $n$ in the sum of less then $\bin(n)$ powers of $2$). Thus, overall we make at most $n-\bin(n) + 1$ queries and the lemma follows.
\end{proof}

Before proceeding with the lower bound we briefly discuss lower bounds that can be obtained by other approaches. It is known that $\spar(\MAJ_n) = 2^{n-1}$~\cite{O'Donnell14book}. Thus from the sparsity lower bound we can only get $\pdec(\MAJ_n) \geq \log \spar(\MAJ_n)/2 = \frac{n-1}{2}$. 

Note also that each input $x \in \zoon$ to $\MAJ_n$ lies in the subcube of dimension at least $\lceil\frac{n-1}{2}\rceil$. Indeed, if $\MAJ_n(x)=1$ just pick a subcube on some subset of variables of size $\lceil\frac{n-1}{2}\rceil$ containing all ones of the input. The case $\MAJ_n(x)=-1$ is symmetrical. Thus, in the approach through certificate complexity we get $\pdec(\MAJ_n) \geq \lceil\frac{n-1}{2}\rceil$.

Finally, we observe that the degree approach also does not give a matching lower bound.

\begin{lemma} \label{lem:maj_degree}
For any $n$ we have $\deg(\MAJ_n)=2^p$ where $p$ is the largest integer such that $2^p \leq n$.
\end{lemma}

The proof of this lemma is provided in Appendix.

It is not hard to see that this lower bound matches the upper bound of Lemma~\ref{lem:maj_upper_bound} only for $n=2^r$ and $n=2^r+1$. On the other hand, for example it is far from optimal by approximately a factor of 2 for $n=2^r-1$ for some $r$.

We next show that Theorem~\ref{thm:main} gives a tight lower bound for parity decision tree complexity of $\MAJ_n$.

\begin{lemma} \label{lem:maj_lower_bound}
$\gran(\MAJ_n) = n - \bin(n)$.
\end{lemma}

\begin{proof}
We will show that $\gran(\MAJ_n) \geq n - \bin(n)$. The inequality in the other direction follows from Lemma~\ref{lem:maj_upper_bound}.

We consider the Fourier coefficient $\wh{\MAJ}_n([n])$ and show that its granularity is at least $n-\bin(n)$. Let $k=\lfloor(n+1)/2\rfloor$. Note that $k$ is  the smallest number such that $\MAJ_n$ is $-1$ on inputs with $k$ ones.

Then we have
\begin{align*}
\wh{\MAJ}_n([n]) &= \frac{1}{2^n}\left(\sum_{i=0}^{k-1} (-1)^i\binom{n}{i} - \sum_{i=k}^{n} (-1)^i\binom{n}{i}\right)\\
&= \frac{1}{2^n}\left(\sum_{i=0}^{n} (-1)^i\binom{n}{i} - 2 \sum_{i=k}^{n} (-1)^i\binom{n}{i}\right) = \frac{1}{2^n} \left(0 - 2 \sum_{i=k}^{n} (-1)^i\binom{n}{i}\right).
\end{align*}
From this we can see that
$$
\gran(\wh{\MAJ}_n([n])) =n - \pow\left(2\sum_{i=k}^{n} (-1)^i\binom{n}{i}\right).
$$

We proceed to simplify the sum of binomials (a very similar analysis is presented in~\cite{SaksW91}):
$$
\sum_{i=k}^{n} (-1)^i\binom{n}{i} = \sum_{i=k}^{n} (-1)^i \left( \binom{n-1}{i-1} + \binom{n-1}{i} \right) = (-1)^{k} \binom{n-1}{k-1}.
$$
Thus it remains to compute $\pow(2\binom{n-1}{k-1})$. For even $n=2h$ we have $k=h$ and $2\binom{n-1}{k-1} = 2\binom{2h-1}{h-1} = \binom{2h}{h}$. For odd $n=2h+1$ we have $k=h+1$ and $2\binom{n-1}{k-1} = 2\binom{2h}{h}$.

By~\cite[Proposition 3.4]{SaksW91} we have $\pow(\binom{2h}{h})=\bin(h)$ (alternatively this can be seen from Kummer's theorem). Finally, notice that $\bin(2h)=\bin(h)$ and $\bin(2h+1)=\bin(h)+1$. It follows that 
$$
\pow\left(2\sum_{i=k}^{n} (-1)^i\binom{n}{i}\right) = \bin(n)
$$
and
$$
\gran(\wh{\MAJ}_n([n])) = n - \bin(n).
$$
\end{proof}

Overall, we have the following theorem.
\begin{theorem}\label{thm:maj}
$$
\pdec(\MAJ_n) = n - \bin(n)+1.
$$
\end{theorem}

As a corollary from this result and Lemma~\ref{lem:mc_vs_pdec} we get the following lower bound on the multiplicative complexity of majority.

\begin{corollary}
$$
\mult(\MAJ_n) \geq n - \bin(n).
$$
\end{corollary}

This improves a lower bound of~\cite{BoyarP08}. Previously our lower bound was known only for $n=2^k$ for some $k$~\cite{BoyarP08}.

\section{Recursive Majority} \label{sec:rmaj}

Next we study the parity decision tree complexity of recursive majority $\RMAJ{k}$. This is a function on $n=3^k$ variables and it can be defined recursively. For $k=1$ we just let $\RMAJ{1}=\MAJ_3$. For $k>1$ we let
$$
\RMAJ{k} = \MAJ_3\left(\RMAJ{k-1},\RMAJ{k-1},\RMAJ{k-1}\right),
$$
where each $\RMAJ{k-1}$ is applied to a separate block of variables.

We start with an upper bound.
\begin{lemma} \label{lem:rmaj_upper_bound}
$\pdec(\RMAJ{k}) \leq (n+1)/2$.
\end{lemma}

\begin{proof}
Basically, recursive majority $\RMAJ{k}$ is a function computed by a Boolean circuit which graph is a complete ternary tree of depth $k$, each internal vertex is labeled by the function $\MAJ_3$ and each leaf is labeled by a (fresh) variable.

To construct an algorithm we first generalize the problem. We consider functions computed by Boolean circuits which graphs are ternary tree, where each non-leaf has fan-in $3$ and is labeled by $\MAJ_3$, and each leaf is labeled by a fresh variable. We will show that if the number of non-leaf variables in the circuit is $l$, then the function can be computed by a parity decision tree of size $l+1$.

The proof is by induction on $l$. If $l=1$, then the function in question is just $\MAJ_3$ and by the results of Section~\ref{sec:maj} it can be computed by a parity decision tree of size $2$.

For the step of induction consider a tree with $l$ non-leaf vertices. Consider a non-leaf vertex of the largest depth. All of its three inputs must be variables, lets denote them by $y$, $z$ and $t$, and in this vertex the function $\MAJ_3(y,z,t)$ is computed. Our first query will be $y\oplus z$. It will tell us whether $y$ and $z$ are equal. If $y=z$ are equal, then $\MAJ_3(y,z,t)=y$, and if $y\neq z$, then $\MAJ_3(y,z,t)=t$. Thus, we can substitute the gate in our vertex by the corresponding variable and reduce the problem to the circuit with $l-1$ non-leaf vertices. By induction hypothesis, the function computed by this circuit can be computed by at most $(l-1)+1=l$ queries. Thus, our original function is computable by $l+1$ queries.

It is left to observe that a complete ternary tree of depth $k$ has $3^{k-1}+\ldots+1=\frac{3^k-1}{2}$ non-leaf vertices and for this tree our algorithm makes $\frac{3^k+1}{2}=\frac{n+1}{2}$ queries.
\end{proof}

Before proceeding to the lower bound we again discuss lower bounds that can be obtained by other techniques. 

First note that each input $x\in \zoon$ lies in the subspace of co-dimension at most $2^k$ on which the function is constant. For this it is enough to show that in each $x$ we can flip $3^k-2^k$ variables without changing the value of the function. This is easy to check by induction on $k$. For $k=1$ there are two variables that are equal to each other and we can flip the third variable without changing the value of the function. For $k>1$ consider inputs to the $\MAJ_3$ at the top of the circuit. Two of them are equal and by induction hypothesis we can flip $3^{k-1}-2^{k-1}$ variables in each of them without changing the value of the function. The last input to the top gate does not affect the value of the function and we can flip all $3^{k-1}$ variables in it. Overall this gives us $3^k-2^k$ variables. This gives us $\cert(\RMAJ{k}) \leq 2^k = n^{\log_3 2}$ which does not give a matching lower bound.

Also note that the polynomial computing $\MAJ_3$ is $p(x)=x_1x_2 \oplus x_2x_3 \oplus x_1 x_3$. The polynomial for $\RMAJ{k}$ can be computed by a simple composition of $p$ with itself. It is easy to see that its degree is $2^k = n^{\log_3 2}$. Thus, an approach through polynomials over $\bb{F}_2$ does not give strong lower bounds.

For Fourier analytic considerations it is convenient to switch to $\moo$ Boolean inputs. For a variable $y\in \zoo$ let us denote by $y'\in \moo$ the variable $y'=1-2y$. For now we will use new variables as inputs to Boolean functions.

The Fourier decomposition of $\MAJ_3$ is
\begin{align} \label{eq:maj_3}
\MAJ_3(y',z',t') = \frac{1}{2}\left(y'+z'+t' - y'z't'\right).
\end{align}

From this the Fourier decomposition of $\RMAJ{k}$ can be obtain by recursion:
\begin{align} \label{eq:rmaj_fourier}
\begin{split}
\RMAJ{k}(x^1,x^2,x^3) = & \frac{1}{2} (\RMAJ{k-1}(x^1)+\RMAJ{k-1}(x^2)+\RMAJ{k-1}(x^3)\\ & - \RMAJ{k-1}(x^1)\cdot\RMAJ{k-1}(x^2)\cdot\RMAJ{k-1}(x^3)),
\end{split}
\end{align}
where $x^1, x^2, x^3$ are blocks of $3^{k-1}$ variables.

Lemma~\ref{lem:spar_lower_bound} can give lower bounds up to $n/2$ and thus in principle might give at least almost matching lower bound. However, this is not the case as we discuss below.

Note that since there is no free coefficient in the polynomial~\eqref{eq:maj_3}, Fourier coefficients arising from all three summands in the right-hand side of~\eqref{eq:rmaj_fourier} will not cancel out with each other: none two of them have equal set of variables. Thus, if we denote $S(k)=\spar(\RMAJ{k})$ we have that $S(1)=4$ and
\begin{equation} \label{eq:rmaj_spar_recursion}
S(k) = 3S(k-1) + S(k-1)^3
\end{equation}
for $k>1$.
On one hand, this means that $S(k) > S(k-1)^3$. This gives $S(k) > 2^{2\cdot 3^{k-1}}$. Thus $\log \spar(\RMAJ{k}) > 2\cdot 3^{k-1}=  2n/3$ and $\pdec(\RMAJ{k})> n/3$.

On the other hand if we let $S'(k)=S(k)+1/2$, it is easy to check that~\eqref{eq:rmaj_spar_recursion} implies
$$
S'(k) < S'(k-1)^3.
$$
Since $S'(1)=9/2$ this gives $S'(k) < 2^{(\log_2\frac{9}{2}) \cdot 3^{k-1}}$. Thus, 
$$
\log \spar(\RMAJ{k}) <  \left(\log_2\frac{9}{2}\right) \cdot \frac{n}{3} < 0.723 \cdot n.
$$ 
Thus Lemma~\ref{lem:spar_lower_bound} can give us a lower bound of at most $0.362\cdot n$. We note that this upper bound on the sparsity can be further improved by letting $S'(k) = S(k)+\alpha$ for smaller $\alpha$.

Now we proceed to the tight lower bound. Again we will estimate $\gran(\wh{\MAJ}^{\otimes k}_{3}[n])$. Observe that this Fourier coefficient can be easily computed from~\eqref{eq:maj_3} and~\eqref{eq:rmaj_fourier}. Indeed, from~\eqref{eq:maj_3} we have that $\left|\wh{\MAJ}^{\otimes 1}_{3}[n]\right| = \frac12$. From~\eqref{eq:rmaj_fourier} we have that
$$
\left|\wh{\MAJ}^{\otimes k}_{3}[n]\right| = \left|\frac{1}{2}(\wh{\MAJ}^{\otimes k-1}_{3}[n])^3\right|.
$$
The numerator of this Fourier coefficient equals to $1$ for any $k$. Thus, denoting $G(n)=\gran(\wh{\MAJ}^{\otimes k}_{3}[n])$ for $n=3^k$ we have $G(3)=1$ and
$$
G(n)=3G\left(\frac{n}{3}\right)+1.
$$
It is straightforward to check that $G(n)=\frac{n-1}{2}$. From this, Theorem~\ref{thm:main} and Lemma~\ref{lem:rmaj_upper_bound} the following theorem follows.

\begin{theorem}
$\pdec(\RMAJ{k}) = \frac{n+1}{2}$, where $n=3^k$ is the number of variables.
\end{theorem}

\section{A Function $f$ with $\pdec(f) > \gran(f)+1$} \label{sec:example}

In this section we provide an example of a function for which our lower bound is not tight. For this we study the family of threshold functions.

For arbitrary $n$ and $k$ we let 
$$
\THR_n^k(x) = -1 \Leftrightarrow \sum_{i=1}^{n} x_i \geq k,
$$
where $x\in \zoon$.
Note that $\MAJ_n = \THR_n^{\lceil n/2\rceil}$.

Our examples will form a subfamily of this family of functions.

To show that our lower bound is not tight we need an approach to prove even better lower bounds. We will do it via the following theorem.

\begin{theorem} \label{thm:hardness_amplification}
For any $s,k, n$ if $\pdec(\THR_n^k)\geq s$, then $\pdec(\THR_{n+2}^{k+1})\geq s+1$.
\end{theorem}

\begin{proof}
We will argue by a contradiction. Assume that $\pdec(\THR_{n+2}^{k+1})\leq s$. We will construct a parity decision tree for $\THR_n^k$ making no more than $s-1$ queries.

Denote the input variables to $\THR_n^k$ by $x=(x_1,\ldots, x_n)$. We introduce one more variable $y$ (which we will fix later) and consider the sequence $x_1,\ldots, x_n, y, \neg y$ as inputs to the algorithm for $\THR_{n+2}^{k+1}$. Note that $\THR_n^k(x)=\THR_{n+2}^{k+1}(x,y,\neg y)$. Our plan is to simulate the algorithm for $\THR_{n+2}^{k+1}$ on $(x,y,\neg y)$ and save one query on our way.

Consider the first query that the algorithm makes to $(x_1,\ldots, x_n, y, \neg y)$. Suppose first that the query does not ask the parity of all variables $(\bigoplus_{i=1}^n x_i) \oplus y \oplus \neg y$ (we will deal with this case later). Since the function $\THR_{n+2}^{k+1}$ is symmetric we can rename the input bits in such a way that the query contains input $y$ and does not contain $\neg y$, that is the query asks the parity $(\bigoplus_{i\in S} x_i) \oplus y$ for some $S\subseteq[n]$. Now it is time for us to fix the value of $y$. We let $y = \bigoplus_{i\in S} x_i$. Then the answer to the first query is $0$, we can skip it and proceed to the second query. For each next query of the algorithm for $\THR_{n+2}^{k+1}$ if it contains $y$ or $\neg y$ (or both) we substitute them by $\bigoplus_{i\in S} x_i$ and $(\bigoplus_{i\in S} x_i)\oplus 1$ respectively. The result is the parity of some variables among $x_1,\ldots, x_n$ and we make this query to our original input $x$. Clearly the answer to the query to $x$ is the same as the answer to the original query to $(x,y,\neg y)$. Thus, making at most $s-1$ queries we reach the leaf of the tree for $\THR_{n+2}^{k+1}$ and thus compute $\THR_{n+2}^{k+1}(x,y,\neg y)=\THR_n^k(x)$.

It remains to consider the case when the first query to $\THR_{n+2}^{k+1}$ is $(\bigoplus_{i=1}^n x_i) \oplus y \oplus \neg y$. This parity is equal to $\bigoplus_{i=1}^n x_i$ and we make this query to $x$. Now we proceed to the second query in the computation of $\THR_{n+2}^{k+1}$ and this query is not equal to $(\bigoplus_{i=1}^n x_i) \oplus y \oplus \neg y$. We perform the same analysis as above for this query: rename the inputs, fix $y$ to the parity of subset of $x$ to make the answer to the query to be equal to 0, simulate further queries to $(x,y,\neg y)$. Again we save one query in this case and compute $\THR_n^k(x)$ in at most $s-1$ queries.
\end{proof}

Next we analyze the decision tree complexity of $\THR_{n}^2$ functions. For them our lower bound is tight, but we need this analysis to use in combination with Theorem~\ref{thm:hardness_amplification} to provide our example.

\begin{lemma} \label{lem:thr_2_complexity}
For even $n$ we have $\pdec(\THR_n^2)=n$ and for odd $n$ we have $\pdec(\THR_n^2)=n-1$.
\end{lemma}
\begin{proof}[Proof sketch]
The proof of the lower bound is technical and is omitted. The complete proof of the lemma can be found in Appendix~\ref{app:thr_2_complexity}.

Here we only prove that the lower bound is tight for odd $n$. To provide an algorithm making at most $n-1$ queries we again will split variables into blocks and again will assume that in the beginning all blocks are of size 1. We split all variables but one into pairs and check whether variables in each pair are equal. After this we have $(n-1)/2$ blocks of size 2 and one block of size 1. If there is a balanced block of size 2, again we can just query one variable from each of the remaining blocks thus learning the number of ones in the input. This allows us to compute the function in at most $n-1$ queries. If all blocks of size 2 contain equal variables, then note that the value of the function does not depend on the variable in the block of size 1. Indeed, $\THR_n^2(x) = 1$ iff $\sum_i x_i \geq 2$ iff there is a block of size 2 containing variables equal to 1. Thus it remains to query one varaible from each block of size 2, which again alows us to compute the function with at most $n-1$ queries.
\end{proof}

We are now ready to proceed to the example of the functions for which the lower bound in Theorem~\ref{thm:main} is tight.

\begin{lemma} \label{lem:thr_3_gran}
For $n=8k+2$ for integer $k$ we have $\gran(\THR_{n}^3) = n-3$.
\end{lemma}

The proof of this lemma is technical and can be found in Appendix~\ref{app:thr_3_gran}.

We now show that for functions in Lemma~\ref{lem:thr_3_gran} their decision tree complexity is greater than their granularity plus one. Note, that since granularity lower bound is not worse than the lower bounds through the sensitivity and the degree, they also do not give tight lower bounds. Also it is easy to see that the approaches through certificate complexity does not give optimal lower bound as well.

\begin{theorem} \label{thm:thr_3}
For $n=8k+2$ for integer $k>0$ we have $\pdec(\THR_{n}^3) = n-1$.
\end{theorem}

\begin{proof}
For the lower bound we note that $n-3$ is odd and thus by Lemma~\ref{lem:thr_2_complexity} we have $\pdec(\THR_{n-2}^2)\geq n-2$. Then by Theorem~\ref{thm:hardness_amplification} we have $\pdec(\THR_{n}^3)\geq n-1$.

For the upper bound we again view the inputs as blocks of size 1 and by checking equality of variables combine all variables but two into blocks of size $4$. If we encounter a balanced block we just query one variable from all remaining block thus learning the number of ones in the input in at most $n-1$ queries. If all blocks contain equal variables, then as in the proof of Lemma~\ref{lem:thr_2_complexity} we observe that two variables outside of blocks of size 4 does not affect the value of the function. Indeed, $\THR_n^3(x) = 1$ iff $\sum_i x_i \geq 3$ iff there is a block of size 4 containing variables equal to 1.
\end{proof}

Thus, we have shown that the lower bound in Theorem~\ref{thm:main} is not tight for $\THR_{8k+2}^{3}$. However, the gap between the lower bound and the actual complexity is 1.

\begin{remark}
We note that from our analysis it is straightforward to determine the complexity of $\THR_{n}^3$ for all $n$. If $n=4k$ or $4k+3$ for some $k$, then $\pdec(THR_n^3)=n$ and if $n=4k+1$ or $n=4k+2$, then $\pdec(THR_n^3)=n-1$. The lower bounds (apart from the case covered by Theorem~\ref{thm:thr_3}) follows from the consideration of $\wh{\THR}^3_n(\emptyset)$ and $\wh{\THR}^3_n([n])$ as in the proof of Lemma~\ref{lem:thr_3_gran}. The upper bound follows the same analysis as in the proof of Theorem~\ref{thm:thr_3}.
\end{remark}

\paragraph*{Acknowledgments} We would like to thank Alexander Kulikov for letting us know about the connection between parity decision trees and multiplicative complexity and for permission to add the proof to the paper. We also would like to thank Alexander for drawing our attention  to the possibility of connection of parity decision tree complexity to the degree over $\bb{F}_2$.

\bibliography{parity}

\section{Appendix: Omitted Proofs}

\subsection{Proof of Lemma~\ref{lem:maj_degree}}

Consider a multilinear polynomial $p$ over $\bb{F}_2$ computing $\MAJ_n$. For a set $S \subseteq [n]$ denote by $c_S$ the coefficient of the monomial $\prod_{i\in S} x_i$ in $p$. Denote $|S|=k$ and denote by $x_S \in \zoon$ the input such that $x_i=1$ iff $i \in S$. By~\cite[Section 2.1]{Jukna12} we have
$$
c_{S} = \bigoplus_{x \leq x_S} \MAJ_n(x),
$$
where the order on $\zoon$ is coordinate-wise.

From this we obtain that
$$
c_S = \sum_{i=\lceil \frac{n}{2}\rceil}^{k} \binom{k}{i} = \sum_{i=\lceil \frac{n}{2}\rceil}^{k} (-1)^i\binom{k}{i} \pmod 2,
$$
where the second equation follows since changing the sign of an integer summand does not change its remainder when divided by 2.

Denote $l = \lceil \frac{n}{2}\rceil$. We can simplify the latter sum as follows:
$$
 \sum_{i=l}^{k} (-1)^i\binom{k}{i} =  \sum_{i=l}^{k} (-1)^i \left( \binom{k-1}{i-1} + \binom{k-1}{i} \right) = (-1)^{l} \binom{k-1}{l-1}.
$$

By Kummer's theorem $\binom{k-1}{l-1}$ is odd iff the summation process of $l-1$ and $k-l$ in binary representation does not have any carry bits. Note that both $l-1=\lceil \frac n2 \rceil-1$ and $k-l\leq \lfloor \frac n2 \rfloor$ are less or equal $n/2$. Thus their binary representations are one bit shorter than the binary representation of $n$. The maximal $k$ for which $\binom{k-1}{l-1}$ is odd (and thus $c_S$ is non-zero) is the one for which $k-l$ has a binary representation inverted compared to $l-1$, that is $(k-l) + (l-1)=k-1$ has a binary representation consisting of ones only. That is, $k$ is a power of 2 not exceeding $n$.

\subsection{Complete Proof of Lemma~\ref{lem:thr_2_complexity}} \label{app:thr_2_complexity}

We start with a lower bound.

Here we will need to consider two Fourier coefficients, $\wh{\THR}^2_n(\emptyset)$ and $\wh{\THR}^2_n([n])$. We start with the latter one.

We have
\begin{align*}
\wh{\THR}_n^2([n]) &= \frac{1}{2^n}\left(\sum_{i=0}^{1} (-1)^i \binom{n}{i} - \sum_{i=2}^{n} (-1)^i\binom{n}{i}\right)\\
&= \frac{1}{2^n}\left(2\sum_{i=0}^{1} (-1)^n\binom{n}{i} -  \sum_{i=0}^{n} (-1)^n\binom{n}{i}\right) = \frac{1}{2^n} \left(2\sum_{i=0}^{1} (-1)^n\binom{n}{i} - 0\right).
\end{align*}
From this we can see that
$
\gran(\wh{\THR}_n^2([n])) =n - \pow\left(\sum_{i=0}^{1} (-1)^n\binom{n}{i}\right) -1
$
and thus
\begin{equation*}
\pdec(\THR_n^2) \geq n - \pow\left(\sum_{i=0}^{1} (-1)^n\binom{n}{i}\right).
\end{equation*}

By the same analysis for $\wh{\THR}^2_n(\emptyset)$ we can show that
$$
\pdec(\THR_n^2) \geq n - \pow\left(\sum_{i=0}^{1} \binom{n}{i}\right).
$$
Note that $\sum_{i=0}^{1} (-1)^n\binom{n}{i} = 1-n$ and $\sum_{i=0}^{1} \binom{n}{i} = 1+n$. From this for even $n$ we clearly obtain a lower bound of $\pdec(\THR_n^2) \geq n$. For odd $n$ it is easy to see that one of the numbers $1-n$ and $1+n$ is not divisible by 4. Thus for odd $n$ we obtain lower bound $\pdec(\THR_n^2) \geq n-1$.

It remains to prove that the lower bound is tight for odd $n$. To provide an algorithm making at most $n-1$ queries we again will split variables into blocks and again will assume that in the beginning all blocks are of size 1. We split all variables but one into pairs and check whether variables in each pair are equal. After this we have $(n-1)/2$ blocks of size 2 and one block of size 1. If there is a balanced block of size 2, again we can just query one variable from each of the remaining blocks thus learning the number of ones in the input. This allows us to compute the function in at most $n-1$ queries. If all blocks of size 2 contain equal variables, then note that the value of the function does not depend on the variable in the block of size 1. Indeed, $\THR_n^2(x) = 1$ iff $\sum_i x_i \geq 2$ iff there is a block of size 2 containing variables equal to 1. Thus it remains to query one varaible from each block of size 2, which again alows us to compute the function with at most $n-1$ queries.

\subsection{Proof of Lemma~\ref{lem:thr_3_gran}} \label{app:thr_3_gran}

For the upper bound we need to consider an arbitrary Fourier coefficient $\wh{\THR}^3_n(S)$.
We have
\begin{align*}
\wh{\THR}_n^3([S]) &= \frac{1}{2^n}\left(\sum_{x, |x|\leq 2} \chi_S(x) - \sum_{x, |x|\geq 3} \chi_S(x)\right)
= \frac{1}{2^n}\left(2\sum_{x, |x|\leq 2} \chi_S(x) -  \sum_{x\in \zoon} \chi_S(x)\right),
\end{align*}
where by $|x|$ we denote $\sum_{i=1}^n x_i$.
The second sum in the last expression is equal to either $2^n$ or $0$ depending on $S$. Thus we have
\begin{equation} \label{eq:gran_thr_3}
\gran(\wh{\THR}_n^3(S)) =n - \pow\left(\sum_{x, |x|\leq 2} \chi_S(x)\right) -1.
\end{equation}

Denote the size of $S$ by $l$. Then we have
$$
\sum_{x, |x|\leq 2} \chi_S(x) = 1 - l + (n - l) + \frac{l(l-1)}{2} - l(n-l) + \frac{(n-l)(n-l-1)}{2},
$$
where the first summand corresponds to $x$ with $|x|=0$, the next two summands correspond to $|x|=1$ and the last three correspond to $|x|=2$.

Rearranging this expression we obtain
$$
\sum_{x, |x|\leq 2} \chi_S(x) = \frac{4l^2 + 2 + (n+1)(n-4l)}{2}.
$$
We need to show that for $n \equiv 2 \pmod 8$ this number is divisible by 4, that is its numerator is divisible by $8$. Since divisibility by $8$ depends only on the remainder of $n$ when divided by $8$, it is enough to check divisibility of the numerator by $8$ for $n=2$.
We have
$$
4l^2 + 2 + (n+1)(n-4l) = 4l^2 + 2 + 3(2-4l) = 4(l^2 -3l + 2),
$$
which is clearly divisible by $8$ for all $l$. Thus $\pow\left(\sum_{x, |x|\leq 2} \chi_S(x)\right)\geq 2$ for $n=8k+2$ and
$$
\gran(\THR_n^3) \leq n - 3.
$$ 

For the lower bound on the granularity it is enough to consider Fourier coefficients $\wh{\THR}^3_n(\emptyset)$ and $\wh{\THR}^3_n([n])$. For them we have
$$
\sum_{x, |x|\leq 2} \chi_\emptyset(x) = 1 + n + \frac{n(n-1)}{2} = \frac{2+n(n+1)}{2} 
$$
and
$$
\sum_{x, |x|\leq 2} \chi_{[n]}(x) = 1 - n + \frac{n(n-1)}{2} = \frac{2 + n(n-3)}{2}.
$$
To show the lower bound it is enough to show that for any $n=8k+2$ at least one of these expressions is not divisible by $8$, that is their numerators are not divisible by $16$. It is straightforward to check that for $n\equiv2 \pmod{16}$ we have $2+n(n+1) \equiv 8 \pmod{16}$ and for $n\equiv10 \pmod{16}$ we have $2 + n(n-3) \equiv 8 \pmod{16}$. In both cases by~\eqref{eq:gran_thr_3} we found a Fourier coefficients with granularity at least $n - 3$.